\documentclass[conference]{IEEEtran}
\IEEEoverridecommandlockouts
\usepackage{cite}
\usepackage{url}
\usepackage{algorithm,amsmath,amssymb,amsfonts}
\usepackage{algorithmic}
\usepackage{graphicx}
\usepackage{textcomp}
\usepackage{xcolor}
\def\BibTeX{{\rm B\kern-.05em{\sc i\kern-.025em b}\kern-.08em
    T\kern-.1667em\lower.7ex\hbox{E}\kern-.125emX}}
\begin{document}

\title{VulSPG: Vulnerability detection based on slice property graph representation learning\\
}

\author{\IEEEauthorblockN{1\textsuperscript{st} Weining Zheng}
\IEEEauthorblockA{\textit{Harbin Institute of Technology} \\
Harbin, China \\
20B903074@stu.hit.edu.cn}
\and
\IEEEauthorblockN{2\textsuperscript{nd} Yuan Jiang}
\IEEEauthorblockA{\textit{Harbin Institute of Technology} \\
Harbin, China \\
jiangyuan@hit.edu.cn}
\and
\IEEEauthorblockN{3\textsuperscript{rd} Xiaohong Su*}
\IEEEauthorblockA{\textit{Harbin Institute of Technology} \\
Harbin, China \\
sxh@hit.edu.cn}
}

\maketitle

\thispagestyle{plain}
\pagestyle{plain}

\begin{abstract}
Vulnerability detection is an important issue in software security. Although various data-driven vulnerability detection methods have been proposed, the task remains challenging since the diversity and complexity of real-world vulnerable code in syntax and semantics make it difficult to extract vulnerable features with regular deep learning models, especially in analyzing a large program. Moreover, the fact that real-world vulnerable codes contain a lot of redundant information unrelated to vulnerabilities will further aggravate the above problem. To mitigate such challenges, we define a novel code representation named Slice Property Graph (SPG), and then propose VulSPG, a new vulnerability detection approach using the improved R-GCN model with triple attention mechanism to identify potential vulnerabilities in SPG. Our approach has at least two advantages over other methods. First, our proposed SPG can reflect the rich semantics and explicit structural information that may be relevance to vulnerabilities, while eliminating as much irrelevant information as possible to reduce the complexity of graph. Second, VulSPG incorporates triple attention mechanism in R-GCNs to achieve more effective learning of vulnerability patterns from SPG. We have extensively evaluated VulSPG on two large-scale datasets with programs from SARD and real-world projects. Experimental results prove the effectiveness and efficiency of VulSPG.
\end{abstract}


\section{Introduction}
Source code-based vulnerability detection is a classic problem, in which considerable efforts have been made in the past decade, such as employing static and dynamic analysis tools to flag potentially dangerous code as candidate vulnerabilities so that it can gain special attention~\cite{lin2020software,votipka2018hackers}. However, many methods rely on rules manually curated by experts in the security domain, which usually suffer from low coverage (leading to false negatives) and high-cost (due to false positives) \cite{engler2001bugs} \cite{RoughAuditing}. Moreover, these methods are not suitable for tracking existing unknown vulnerabilities on a large scale~\cite{zhou2017automated}. 

Along with the rapid development of artificial intelligence and the increasing accessibility of large scale open-source code repository, many works begin to apply deep learning to vulnerability detection considering that deep neural networks can extract features in an automatic way, instead of being designed by domain experts. For example, Li et al. \cite{li2021sysevr} proposed a systematic framework SySeVR to represent the syntax and semantic information pertinent to vulnerabilities, in which program slicing technique \cite{kilpatrick1988program} is used to produce smaller code segments (i.e., some possibly discontinuous but semantic and vulnerability-related statements), making it applicable for deep learning networks. Although SySeVR improves the performance of deep learning models in vulnerability detection, it still has two weaknesses: (1) Code segments generated by SySeVR can only cover  limited types of vulnerabilities, especially for real-world programs. (2) A sequential neural work (e.g., LSTM or GRU) is used to encode code segments in the form of plain texts, which may lose the information of data and control dependencies among statements in code segments, and therefore cannot effectively learn the non-sequential features of graph-based information. 

To preserve the complex structral and sementic information of source code and effectively learn these non-sequential features of graph-based information, Zhou et al. \cite{zhou2019devign} proposed a GGNN-based  vulnerability detection model Devign, which first integrates syntax and dependency information of source code into a joint graph representation, and then extracts vulnerability-relevant features from the composite graph with the help of GNNs. However, due to the very small proportion of vulnerable statements in real-world functions, the large amount of information irrelevant with any vulnerability patterns in the composite graph will hinder the accurate learning of the vulnerability patterns.

To mitigate the above problems, we define a novel graph-and-slice based code representation named Slice Property Graph (SPG). Compared with other graph-based code representations, the proposed SPG differs in that it aims to not only preserve the rich semantics and explicit structural information that may be relevant to vulnerabilities, but also eliminate irrelevant information as much as possible to reduce the complexity of graphs. To this end, based on the four SyVCs (Syntax-based Vulnerability Candidates) proposed by Li et al. \cite{li2021sysevr}, we first add two kinds of new SyVCs to improve the vulnerability coverage of program slicing. Using these six kinds of SyVCs as slicing criterion, we use program slicing technique to extract the slice nodes which might be relevant to vulnerabilities. Then, we take the data dependency, control dependency, and function call dependency between the extracted slice nodes as edges to generate SPG. All these dependency are extracted from the code property graph (i.e. CPG) \cite{yamaguchi2014modeling} which combines abstract syntax tree (AST) \cite{noonan1985algorithm}, control flow graph (CFG) \cite{ferrante1987program} and program dependency graph (PDG) \cite{ferrante1987program} into a single data structure. Next, we propose a new vulnerability detection approach, i.e., VulSPG, which combines the improved relational graph convolutional network (R-GCN) \cite{schlichtkrull2018modeling} model with triple attention mechanism to identify potential vulnerabilities in SPGs. More concretely, the R-GCN can perform representation learning on SPGs with multiple attributes and dependencies, and the triple attention mechanism helps the detection model focus on vulnerability features and suppress unimportant ones via learning attention weight. We evaluated the proposed method on two datasets, and the experimental results demonstrate that VulSPG can identify vulnerabilities more accurately than the baselines.

The main contributions of our work are:
\begin{itemize}
\item We design a new code representation SPG, which can encode vulnerable code into a graph data structure and eliminate as much redundant information that may not be relevant to vulnerabilities as possible. To improve the vulnerability coverage of SPG, two new kinds of SyVCs are added to generate SPGs.  
\item We propose the R-GCN model with triple attention mechanism for representation learning of SPG, which can efficiently and accurately capture vulnerability-related features in SPG for vulnerability detection tasks.
\item We conduct intensive evaluation of VulSPG to verify the effectiveness and superiority of the proposed method through comparative experiments with other methods. The data is available at https://github.com/zhengweining1/VulSPG.
\end{itemize}
\section{Motivation}

\subsection{Motivation for Slice Property Graph}\label{IIA}

\begin{figure}[b]
\vspace{-0.6cm}
\centerline{\includegraphics[width=9cm, height=5.5cm]{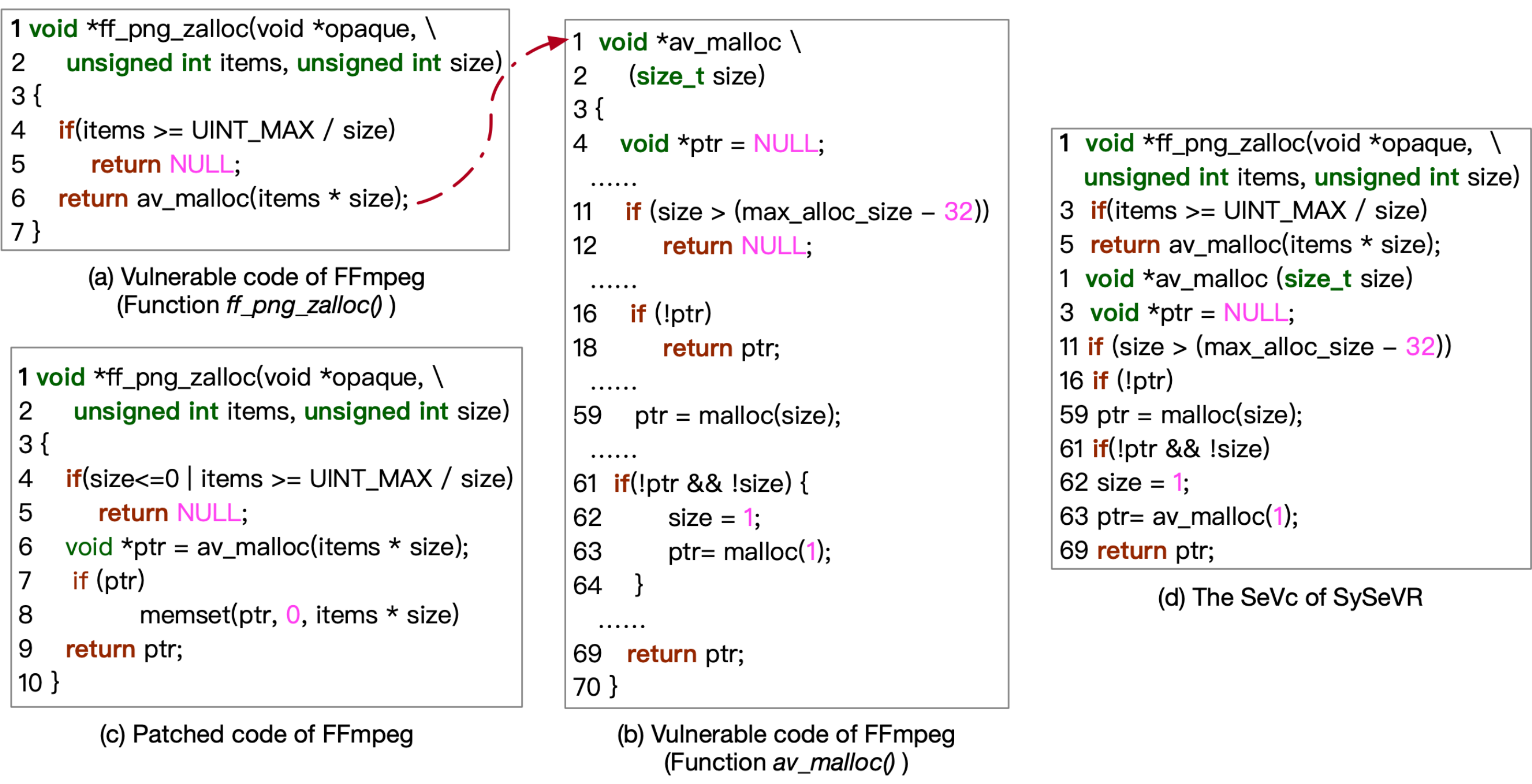}}
\vspace{-0.3cm}
\caption{Motivation example.}
\label{fig1}
\end{figure}

Fig.~\ref{fig1}(a) and Fig.~\ref{fig1}(b) shows the vulnerable code in the FFmpeg project that was fixed in revision \emph{486f0b0}, which contains two vulnerabilities: a) Divide By Zero. When the program passes the parameter \emph{size} to the function \emph{ff\_png\_zallo()}, an error is raised at the statement \emph{if(items $\ge$ UINT\_MAX / size)} if the attacker sets the value of \emph{size} to 0. b) Uninitialized Memory. The function \emph{ff\_png\_zallo()} calls the \emph{av\_malloc()} to request memory without initializing it. Fig.~\ref{fig1}(c) shows the patched code for this vulnerable program.

From this example, we obtain the following  observations:

\textbf{Observation 1.} Vulnerabilities in real-world projects usually span multiple functions or files. For example, the Uninitialized-Memory vulnerability in Fig.~\ref{fig1} spans the function \emph{ff\_png\_zallo()} and \emph{av\_malloc()}. Therefore, interprocedural analysis is necessary for detecting vulnerabilities across multiple functions or even files, that is, calling relations between functions should be included in code representation to enable it to entirely characterize these vulnerabilities.  

\textbf{Observation 2.} There are usually a large number of vulnerability independent statements in vulnerable code. For example, the vulnerable code in Fig.~\ref{fig1} has 70 lines of statements, but only a few of them are logically and semantically to any potential vulnerability. The divide-by-zero vulnerability is only related to the parameter \emph{size} and the conditional statement \emph{if(items$\ge$UINT\_MAX/size)}. The uninitialized-memory vulnerability is related to \emph{return av\_malloc(items*size);}, \emph{ptr = malloc(size);}, \emph{return ptr;}. Taking the vulnerable code shown in Fig.~\ref{fig1} as example, the composite graph proposed by Devign \cite{zhou2019devign} can comprehensively characterize the program by merging AST, CFG, Data flow Graph (DFG) and Natural Code Sequence (NCS) into one graph. It contains 171 nodes and 322 edges, but only several nodes and edges are relevant to the vulnerability. 

SySeVR \cite{li2021sysevr} proposed by Li et al. utilizes program slicing technique to generate a set of code statements that are semantically related to the vulnerability called SeVCs (Semantic-based Vulnerability Candidates). In fact, SeVCs is an ordered set of statements which include backward, forward and interprocedural program slice. Fig.~\ref{fig1}(d) shows the SeVC generated by SySeVR based on the sensitive function \emph{malloc()} (a FC-kind SyVC). It provides an entire representation of the uninitialized-memory vulnerability through interprocedural slicing, and reduces a large number of statements that are unrelated to the vulnerability. However, SySeVR leverages token-based sequential representation, which cannot reflect complex structural information and dependencies in the code. The need to model the logic and structure of programs in vulnerability detection tasks has been demonstrated in \cite{zhou2019devign} and \cite{wang2020combining}. Furthermore, the SeVCs generated by using 4 types of slicing criterions can only cover part of the vulnerable statements in the source code. Taking the SARD dataset provided by Li et al. themselves as an example, about 93.6\% of all vulnerable codes in the dataset can be covered by SeVCs, but there are still 890 vulnerable codes that cannot be covered.

To overcome these challenges, we are encouraged to design a more effective code representation, which needs to satisfy several conditions. 1) It should contain the calling relations between functions. 2) It is capable of eliminating redundant information that may not be relevant to the vulnerability as much as possible but can cover more vulnerable code. 3) It is able to characterize the logic and structure of the vulnerable code with a small yet semantically rich graph. To this end, we design SPG, a edge-labeled and directed attributed multigraph, for code representation. The definition and the generation process of SPG can be found in section \ref{III} and section \ref{IVA}.

\subsection{Motivation for Using Graph Neural Network}

Current popular approaches convert the various representations of code into a tiled sequence and then use deep learning models (usually CNNs and RNNs) to learn the code representation as natural language sequences \cite{lee2017learning} \cite{russell2018automated} \cite{choi2017end}. However, SPG as a kind of graph data is irregular, each node in the graph may have a different number of neighboring nodes and the data samples of each node are not independent of each other but are related to other nodes, making it hard for general deep learning algorithms to learn the structural features of the graph. Hence, to obtain a more accurate characterization of the vulnerability code, we select Graph Neural Networks (GNNs) to learn the topology structure and node representation information directly from the SPG. We choose Relational Graph Convolutional Networks (R-GCNs) rather than other graph convolutional networks \cite{duvenaud2015convolutional}, such as GraphSAGE \cite{hamilton2017inductive}, Gated graph sequence neural networks (GGNN) \cite{DBLP:journals/corr/LiTBZ15}, to learn the characteristics of nodes for two reasons. First, the edges of the SPG are directed, and the directional information of the edges cannot be properly learned by using non-directed GNN models. Second, SPG contains multiple relationships between its nodes, and R-GCN can better exploit these relationships to learn representations of source code.

\subsection{Motivation for Using the Triple Attention Module}

Different syntax elements in the source code are of different importance for the detection of vulnerabilities. For example, a statement of type CallExpression related to an API function call may contain more information about the vulnerability. Besides, the code structure contains several other independent dependencies, such as data dependencies, control dependencies and function call dependencies, and the importance of these dependencies varies from one vulnerability to another. For example, data dependencies may be more important than control dependencies in vulnerabilities related to the use of arrays. Therefore, we introduce an attention mechanism to pay more attention to the key vulnerability features so that these features have a greater impact on the classification results. However, there are still some obstacles in applying them to vulnerability detection. First, the attention mechanism should fit both the proposed SPG and the GNN model used in this paper. Second, the attention mechanism needs to be better adapted to the problem of vulnerability detection.


\section{OVERVIEW}\label{III}

\begin{figure*}[htbp]
\centerline{\includegraphics[width=17cm, height=9.5cm]{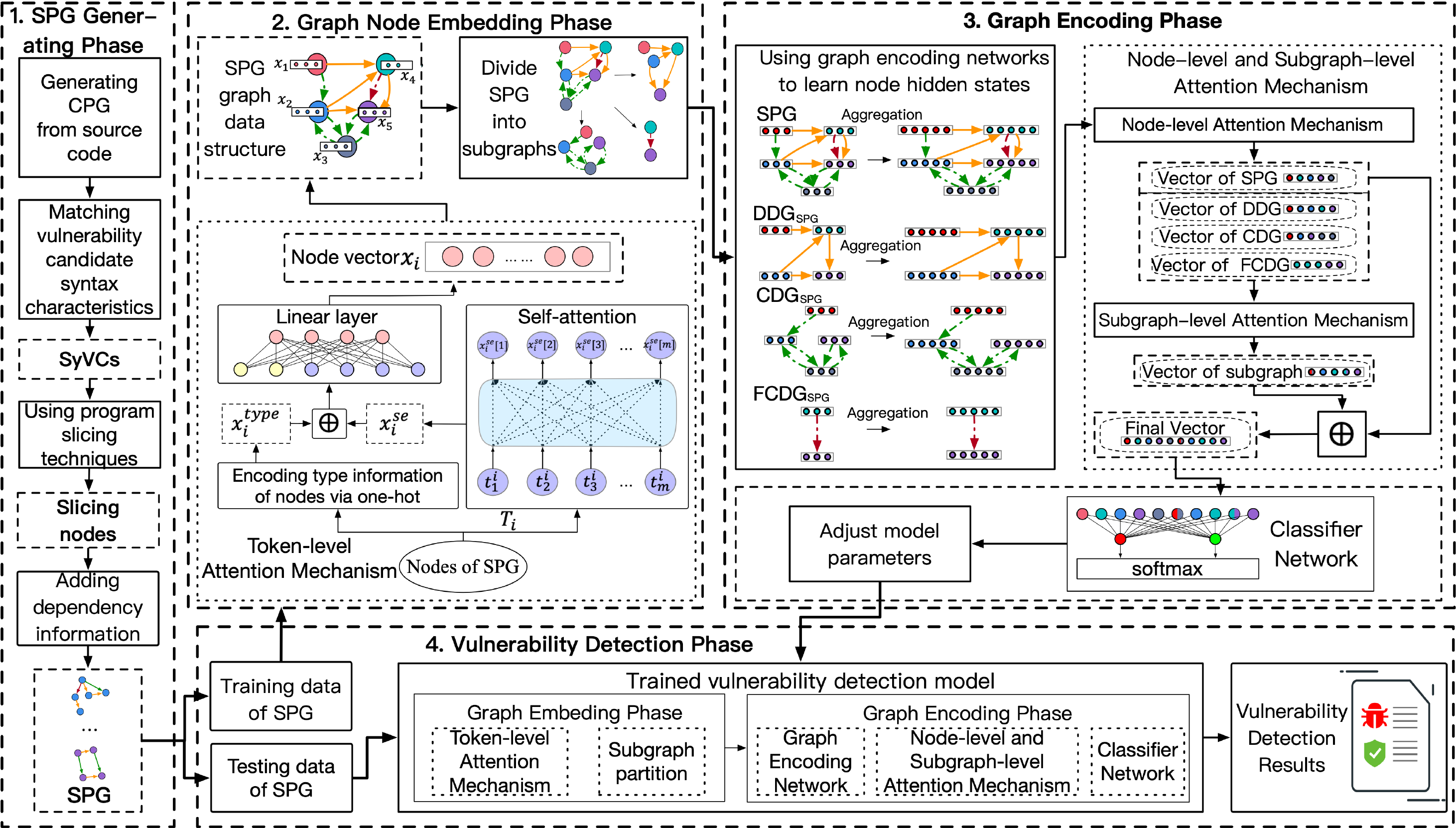}}
\vspace{-0.3cm}
\caption{The processing framework of VulSPG.}
\label{fig2}
\vspace{-0.6cm}
\end{figure*}

As shown in Fig.~\ref{fig2}, the proposed VulSPG method contains four phases: 

1) \emph{SPG Generating Phase.} To generate the corresponding SPGs of programs, we first traverse  the program's AST and identify key points (or called SyVCs~\cite{li2021sysevr}) with special syntactic features that are likely to trigger any vulnerability. Then, program slicing technique is applied to find the semantically related nodes (called slice nodes) in PDG to SyVCs.
Finally, the slice nodes are connected, through the data-, control- and function call-dependency, to form the SPG structures as shown in Fig. \ref{fig4}. For clarity, the proposed SPGs is defined as follows:

\textbf{Slice Property Graph (SPG)}: SPG is a novel code representation which merges data dependency graph (DDG), control dependency graph (CDG) and function call dependency graph (FCDG) into a joint graph data structure. Formally, a slice property graph can be denoted as $ g^{SPG}=(V,E,X,D) $ where $V$ is a set of slice nodes, $E$ is a set of directed edges, $X$ is the set of property values of nodes, and $D$ is the set of property values of directed edges. 


2) \emph{Graph Node Embedding Phase.} In this phase, we initialize the feature vector for each node in SPG. Firstly, every node of SPG described one code statement is lexied into a sequence of tokens, each of which is then embedding into a continuous distributed vector via the pre-trained word2vec model.
Secondly, using these vectors as input,
the semantic-feature vectors of the nodes in SPG are obtained by the token-level attention method in the triple attention mechanism. Next, the node type information is embedded into the node-type vector via one-hot encoding, and it is fed into a linear layer by concatenating it with the semantic-feature vector. The output result is the initial vector of SPG nodes. 

3) \emph{Graph Encoding Phase.} In this phase, we first learn the hidden states of the nodes of SPG, DDG, CDG, and FCDG through graph encoding networks, respectively. Then, the hidden states of the nodes can be used for weighted summation to calculate the features of SPG and its subgraphs including DDG, CDG, and FCDG. The weights can be obtained through the node-level attention mechanism. Next, we obtain the feature vector of the subgraph by integrating all subgraphs through a subgraph-level attention mechanism. Finally, the SPG feature vector and the subgraph feature vectors are concatenated and fed into the classifier network to train the model.

4) \emph{Vulnerability detection phase.} In this phase, we use the model trained in the previous phases to detect whether the input SPG is vulnerable or not.

\section{METHODOLOGY}

\subsection{SPG Generating Phase}\label{IVA}

In this section, we describe the process of generating the SPG which can be divided into three steps.

\textbf{Step 1.} The source code is first parsed by using open-source tool joern\footnote{https://joern.readthedocs.io/} to produce a series of code representations (e.g., AST, PDG, CPG), which enable us to analyze the structure of source code and function calls as well as dependencies in the program. 

\begin{figure}[b]
\vspace{-0.6cm}
\centerline{\includegraphics[width=9cm, height=5.5cm]{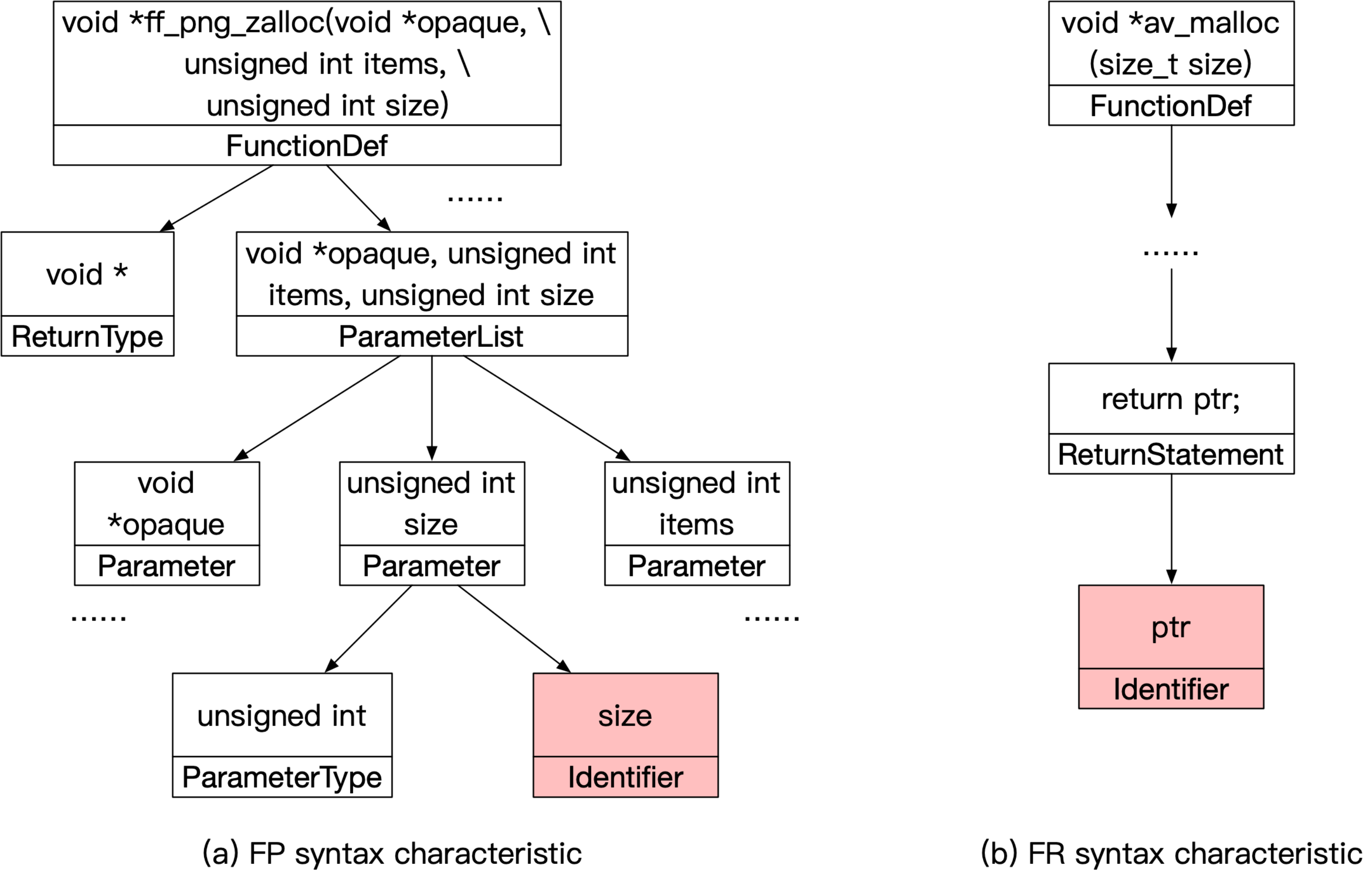}}
\caption{Examples for FP-kind and FR-kind SyVCs}
\label{fig3}
\vspace{-0.3cm}
\end{figure}

\textbf{Step 2.} We search ASTs for SyVCs as a set of slicing criterions through matching some vulnerability candidate syntax characteristics. The paper uses following six types of SyVCs: API/Library Function Call (FC), Array Usage (AU), Pointer Usage (PU), Arithmetic Expression (AE), Function Parameter (FP) and Function Return statement (FR), where the first four types come from the work of Li et al \cite{li2021sysevr} (please refer their work for details), and the other two types are proposed in this paper to improve the coverage of the vulnerability by the SPG. Fig. \ref{fig3} shows an example of matching the FP-kind and FR-kind syntax characteristics for the vulnerable code given in Fig. \ref{fig1}. We use this example to illustrate how to match FR-kind and FP-kind SyVCs. 
\begin{itemize}
\item The highlighted node in Figure 3(a) matches the FP-kind syntax characteristic, because it satisfies the following two conditions: 1) its type is ``Identifier", and 2) it has a parent node of type ``Parameter". 
\item The highlighted node in Figure 3(b) matches the FR-kind syntax characteristic, because it satisfies the following two conditions:1) its type is ``Identifier", and 2) it has a parent node of type ``Returnstatement".
\end{itemize}

\textbf{Step 3.} Using the SyVCs obtained in step 2 as the slicing criterion, we traverse the PDG obtained in step 1 to get a set of statements (called slice nodes) related to each slicing criterion. Next, we obtain data- and control-dependencies as well as function calls among slice nodes from CPG. Finally, the union of all dependency sets and functions calls induces the proposed SPG, where nodes are slice nodes and directed edges represent the dependency and call relationships between slice nodes. The detailed process is shown in Algorithm 1, which can be divided into four parts:

\begin{figure}[htbp]
\vspace{-0.3cm}
		\label{alg2}
		\renewcommand{\algorithmicrequire}{\textbf{Input:}}
		\renewcommand{\algorithmicensure}{\textbf{Output:}}
		\begin{algorithm}[H]
		    \footnotesize
			\caption{Generating SPGs by SyVCs}
			\begin{algorithmic}[1]
				\REQUIRE A program $P=\{f_1,...,f_\eta\}$, a set $Y$ of SyVCs;    
				\ENSURE The set $G$ of SPGs 
				\STATE $G \leftarrow \emptyset$;
				\STATE Generate a CPG $g^{CPG}$ for $P$;
				\FOR {each function $f_i\in P$}
				\STATE Generate a PDG $g_i^{PDG}$ for $f_i$;
				\ENDFOR
				\FOR {each $ce_j\in Y$ in $g_i^{PDG}$}
				\STATE $SN(ce_j),FSN(ce_j),IFSN(ce_j),BSN(ce_j),IBSN(ce_j)\leftarrow \emptyset$;
				\STATE Generate the set $FSN(ce_j)$ of forward slice nodes from $g_i^{PDG}$;
				\FOR {each node $n_k\in FSN(ce_j)$}
				\IF{$n_k$ or its child node is of type ``CallExpression" \& $n_k$ contains the function name $f_l\in P$ called by $f_i$}
				\STATE Generate the set $IFSN_k(ce_j)$ of interprocedural forward slice nodes from $g_l^{PDG}$;
			    \ENDIF
			    \STATE $IFSN(ce_j)\leftarrow IFSN(ce_j)\cup IFSN_k(ce_j)$;
			    \ENDFOR
			    \STATE Generate the set $BSN(ce_j)$ of backward slice nodes from $g_i^{PDG}$;
			    \FOR{each node $n_z\in g^{CPG}$}
			    \IF{$n_z$ or its child node is of type ``CallExpression" \& $n_z$ contains the name of function $f_i$ \& $n_z$ belongs to the function $f_m\in P$ calling $f_i$ }
			    \STATE Generate the set $IBSN_z(ce_j)$ of interprocedural backward slice nodes from $g_m^{PDG}$;
			    \ELSIF{$n_z$ is a node of type ``ReturnStatement" \& $n_z$ belongs to the function $f_l\in P$ called by $f_i$ }
			    \STATE Generate the set $IBSN_z(ce_j)$ of interprocedural backward slice nodes from $g_l^{PDG}$;
			    \ENDIF
			    \STATE $IBSN(ce_j)\leftarrow IBSN(ce_j)\cup IBSN_z(ce_j)$;
			    \ENDFOR
			    \STATE $SN(ce_j)\leftarrow FSN(ce_j)\cup IFSN(ce_j)\cup BSN(ce_j)\cup IBSN(ce_j)$
			    \FOR{each edge $e_q\in g^{CPG}$}
			    \IF{$src(e_q)\in SN(ce_q)$ \& $end(e_q)\in SN(ce_q)$ \& $e_q$ is data/control dependency edge} 
			    \STATE Add nodes $src(e_q),end(e_q)$ and edge $e_q$ to $g_j^{SPG}$;
			    \ENDIF
				\ENDFOR
				\STATE Add function call dependency edges to $g_j^{SPG}$;
				\STATE $G\leftarrow G\cup g_j^{SPG}$
				\ENDFOR
				\STATE Return $G$;
			\end{algorithmic}
		\end{algorithm}
	\vspace{-0.9cm}
	\end{figure}
1) \textbf{Generating the CPG and PDGs (Line 1-5 in Algorithm 1)}: We first generate the CPG $g^{CPG}$ for program $P$. Then, we generate a PDG $g_i^{PDG}$ for each function $f_i\in P$. 

2) \textbf{Extracting the set of forward slice nodes (Line 6-14 in Algorithm 1)}: First, using one SyVC $ce_j$ as a slicing criterion, we forward traverse $g_i^{PDG}$ of function $f_i$ to obtain the set $FSN(ce_j)$ of forward slice nodes. Then, for each node $n_k$ in $FSN(ce_j)$, it is considered as caller nodes if it or its child is of type ``CallExpression". For the corresponding function $f_l$, we perform forward traverse over the PDG $g_l^{PDG}$
to obtain the set $IFSN_k(ce_j)$ of interprocedural forward slice nodes. Finally, we merge all the interprocedural forward slice nodes to obtain the set $IFSN(ce_j)$ of interprocedural forward slice nodes.

3) \textbf{Extracting the set of backward slice nodes (Line 15-23 in Algorithm 1)}: First, using SyVC $ce_j$ as slicing criterion, we backword traverse $g_i^{PDG}$ of function $f_i$ to obtain the set $BSN(ce_j)$ of backward slice nodes. Next, we query all nodes in $g^{CPG}$ to search a node $n_z$ that satisfies one of the following two conditions: 1) $n_z$ or its child node is of type ``CallExpression" and calls $f_i$; 2) $n_z$ is of type ``ReturnStatement" and belongs to the function $f_l$ called by $f_i$; Finally, we backward traverse the PDG of the corresponding function using $n_z$ as a starting node to obtain the set $IBSN_z(ce_j)$ of interprocedural backward slice nodes and merge all the interprocedural backward slice nodes to obtain the set $IBSN(ce_j)$ of interprocedural backward slice nodes. 

4) \textbf{Generating the SPG (Line 24-33 in Algorithm 1)}: First, we merge the sets $FSN(ce_j)$ and $IFSN(ce_j)$ obtained from part 2) and the sets $BSN(ce_j)$ and $IBSN(ce_j)$ obtained from part 3) to generate the set $SN(ce_j)$ of slice nodes. Next, we add each control or data dependency edge $e_q$, the source node $src(e_q)$ and the end node $end(e_q)$ of $e_q$ in $g^{CPG}$ to SPG $g_j^{SPG}$ if $src(e_q)\in SN(ce_q)$ and $end(e_q)\in SN(ce_q)$. Then, we analyze the calling relationships between functions according to $g^{
CPG}$ and add the function call dependency egdes to $g_j^{SPG}$. Specifically, if function $f_1$ calls function $f_2$, we add function call dependency edge with node in $f_1$ which calls function $f_2$ as the source node and ``FunctionDef" node in $f_2$ as the end node. 

\begin{figure}[b]
\vspace{-0.6cm}
\centerline{\includegraphics[width=9cm, height=4.5cm]{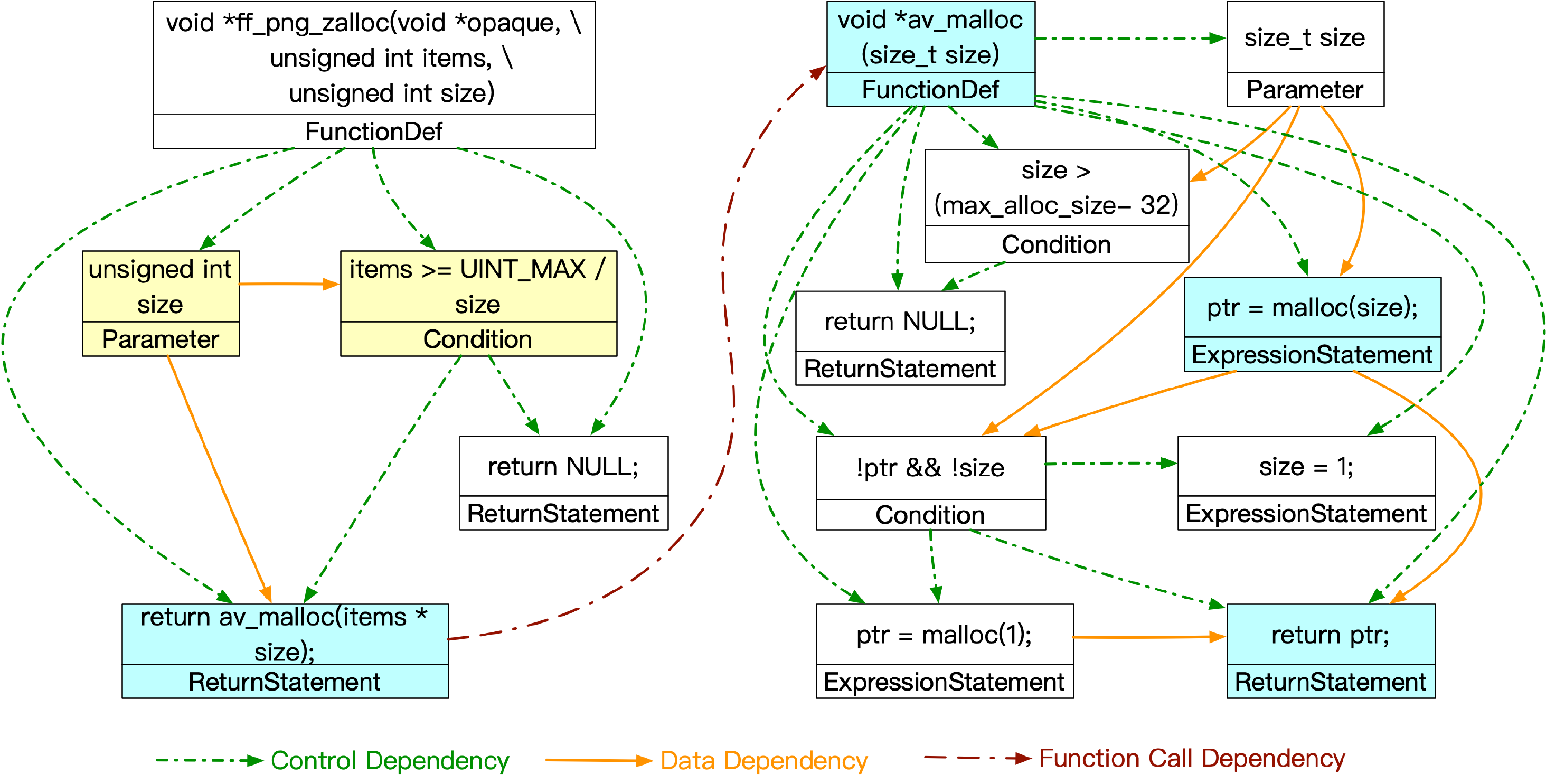}}
\caption{An example of SPG for the vulnerable code in Fig. \ref{fig1}}
\vspace{-0.3cm}
\label{fig4}
\end{figure}

Considering the tradeoff between time cost and performance, only one layer depth of call function is considered in our inter-procedural analysis. Taking the vulnerable code in Fig. 1 as an example, we use the FP-kind SyVC \emph{size} as a slicing criterion to generate the SPG shown in Fig. \ref{fig4}. The yellow nodes which connected by a data dependency edge in SPG are related to the divide by zero vulnerability, illustrating that the vulnerability is only related to data dependency. The blue nodes are related to the uninitialized memory vulnerability. These four nodes are connected in turn by function call dependency edge, control dependency edge, and data dependency edge, illustrating the necessity of representing the vulnerability by these three types of dependencies together.

\subsection{Graph Node Embedding Phrase}\label{IVB}
\textbf{Embedding Nodes of SPG:} To encode the semantic and type information of nodes in the SPG, we use the following to generate the initial embedded vectors of the SPG nodes. For the semantic information of the nodes, we first perform lexical analysis for each statement and create a vector of lexical tokens via the pretrained word2vec model. Next, the dependencies between tokens in each statement are captured using a token-level attention mechanism to obtain a semantic feature vector of the SPG node, the detailed procedure of which can be found in Section \ref{IVD}. For the type information of nodes, we represent the type of each node as a one-hot vector. Since the semantic feature vector and type feature vector do not belong to the same semantic space, we concatenate the two vectors and feed them to a linear layer to finally obtain the initial embedded vector of the node. The parameters in the linear layer can be trained simultaneously with the vulnerability detection model. Formally, a SPG $g^{SPG}(V,E,X,D)$ is a directed multigraph with labeled edges. For each node $v_i\in V$ in SPG, there is a corresponding $x_i\in X$ to represent its feature vector, which is obtained from the semantic feature vector and the type feature vector through a linear layer, i.e. $x_i=W_l(x_i^{se}||x_i^{type})+b_l$. $W_l$ and $b_l$ are the parameters that can be learned in the linear layer, $x_i^{se}$ and $x_i^{type}$ are respectively the corresponding semantic feature vector and the type feature vector of node $v_i$. $x_i$ is the corresponding initial embedded vector of node $v_i$, the symbol $||$ represents the concatenating operation. For each edge $e_j\in E$ in SPG, a corresponding $d_j\in D$ denotes its label, where D contains three types: data- , control-, and function call-dependency.

\textbf{Dividing subgraphs of SPG:} To better distinguish dependencies in SPG and provide a basis for the attention mechanism based on subgraphs, we divide SPG into three types of subgraphs according to the dependencies in SPG, including Control Dependency Graph ($CDG_{SPG}$), a Data Dependency Graph ($DDG_{SPG}$) and a Function Call Dependency Graph ($FCDG_{SPG}$). If the program does not contain a certain dependency that causes the subgraph to be empty, it can be directly treated as a 0-vector of the same dimension.

\subsection{Graph Encoding Network}\label{IVC}

In this section, we use R-GCN to perform representation learning on SPG and its subgraphs. A graph data structure can be denoted as $g^G(V,E,X,D)$, where the superscript $G\in \{SPG,CDG_{SPG},DDG_{SPG},FCDG_{SPG}\}$ refers to the type related to the graph \textit{g}. For each graph, we can use a R-GCN to learn the hidden state of each node in each layer as follows.
\vspace{-0.1cm}
\begin{equation}
\vspace{-0.2cm}
h_i^{(l)}=\sigma(\sum_{d\in D}\sum_{v_j\in N_i^d}\frac{1}{\vert N_i^d\vert}h_j^{(l-1)}W_d^{(l-1)}+h_i^{(l-1)}W_0^{(l-1)})
\label{eq1}
\end{equation}
where $l<L$ denotes the $l$-th layer. $h_i^{(l)}\in R^z$ is the hidden state of the node $v_i$ at layer $l$, and $h_i^{(l-1)}\in R^z$ is the hidden state of the node $v_i$ at the previous layer $l-1$, where both $h_i^{(l)}$ and $h_i^{(l-1)}$ are \textit{z}-dimensional vectors. $d\in D$ denotes an edge type that could be one of the dependency relationships $\{DATA,CONTROL,FUNCTION\_CALL\}$. $N_i^d$ denotes the set of neighboring nodes that point to node $v_i$ under dependency $d$, i.e., if $v_j\in N_i^d$, then there must exist an edge with type $d$ from $v_j$ to $v_i$. $\vert N_i^d\vert$ denotes the number of nodes in the set $N_i^d$. Both $W_d^{(l-1)}\in R^{z\times z}$ and $W_0^{(l-1)}\in R^{z\times z}$ are weight matrices, but $W_d^{(l-1)}$ corresponds to the dependency $d$ while $W_0^{(l-1)}$ has no corresponding dependency. The node initialization state can be seen as the hidden state in the network at layer 0, i.e., $h_i^{(0)}=x_i\in R^z$. At this stage, the weight matrix will also turn to be $W_d^{(0)}$, $W_0^{(0)}\in R^{z\times z}$. After performing the aggregation operation on $g^G$ with $L$ layers of R-GCN, we concatenate the hidden states of the nodes in each layer as the final node representation, i.e., $h_i^{G}=(h_i^{(0)}||h_i^{(1)}||\cdots ||h_i^{(L)})\in R^{(L+1)\cdot z}$.

\subsection{The Triple Attention Mechanism}\label{IVD}


\textbf{Token-level Attention mechanism:} The token-level attention mechanism is primarily designed to obtain dependency information between tokens within a statement to help generate a more accurate semantic vector representation of each statement node. Specifically, the tokens in each statement node are first embedded into fixed-dimensional vectors via a pre-trained word2vec model. Next, we add positional embeddings to the token vectors to encode the position of the tokens with repspect to the sequence. Finally, these vectors are fed into the attention layer to learn the semantic vector representation of each statement node. 
The reason for using multi-head self-attention rather than typical RNNs (GRU, LSTM, et al.) or CNN is as follows: (1) the self-attention method uses a large number of parallel operations and is computationally faster than most of the other techniques mentioned above. (2) compared to RNN and CNN, the self-attention method has been shown to be more capable to learn long-range dependencies among tokens \cite{vaswani2017attention}.

Formally, let the statement node $v_i$ contains $m$ tokens and $T_i=[t_{i,1},t_{i,2},\cdots,t_{i,m}]^T\in R^{m\times c}$ be the semantic matrix corresponding to $v_i$, where $t_{i,j}\in R^c$ denotes the embedding vector of the $j$-th token in the statement. Note that the positional encodings used in $t_{i,j}$ are calculated in the same way as taken in \cite{vaswani2017attention}. Then, the semantic vector $x_i^{se}$ corresponding to the node $v_i$ can be calculated in (\ref{eq2}).
\vspace{-0.25cm}
\begin{equation}
\begin{aligned}
&Attention(Q,K,V)=softmax(\frac{QK^T}{\sqrt{c}})V\\
&MutiHead(Q,K,V)=(head_1||head_2||\cdots||head_a)\\
&where\ head_p=Attention(QW_p^Q,KW_p^K,VW_p^V)\\
&x_i^{se}=flatten(MutiHead(T_i,T_i,T_i))
\end{aligned}
\label{eq2}
\end{equation}
where $MutiHead$ is a multi-head self-attention method that maps a variable-length sequence of code representations to another sequence of equal length, which contains the semantic features of the tokens in the statement. $flatten$ is a function that flattens the resulting semantic sequence into a one-dimensional vector for easy computation of node initialization later (Section \ref{IVB}).

\textbf{Node-level Attention mechanism:} We establish the node-level attention mechanism to obtain the vector representations of SPG and its three subgraphs, as shown in the right part of Fig. \ref{fig2}. For the SPG and its subgraphs, we refer to the attention method in SAGPOOL \cite{lee2019self} to calculate the attention score of each node by taking into account both the characteristics of the nodes themselves and the topology of the graph. The calculation equation is as follows:
\vspace{-0.2cm}
\begin{equation}
\begin{aligned}
&z_i^{G}=\sigma(\sum_{d\in D}\sum_{v_j\in N_i^d}\frac{1}{\vert N_i^d\vert}h_j^{G}\Theta_d^{G}+h_i^{G}\Theta_0^{G})\\
&where\ G\in \{SPG,CDG_{SPG},DDG_{SPG},FCDG_{SPG}\}
\end{aligned}\label{eq3}
\vspace{-0.3cm}
\end{equation}
where $z_i^{G}\in R$ denotes the attention score of the node $v_i$ in the graph $g^G$. $\Theta_d^{G}\in R^{(L+1)\cdot z}$ and $\Theta_0^{G}\in R^{(L+1)\cdot z}$ are both learnable parameters in the model, $\Theta_d^{G}$ corresponds to the dependency $d$ while $\Theta_0^{G}$ has no corresponding dependency. The other parameters are the same as those presented in Section \ref{IVC}. 
\vspace{-0.35cm}
\begin{equation}
\vspace{-0.2cm}
\begin{aligned}
&a_i^{G}=softmax(z_i^{G})=\frac{exp(z_i^{G})}{\sum_{k\in \vert V_{G}\vert}exp(z_k^{G})}\\
&S_{G}=\sum_{i\in \vert V_{G}\vert}\alpha_i^{G}\cdot h_i^{G}
\end{aligned}\label{eq4}
\vspace{-0.1cm}
\end{equation}

To avoid loss of information, instead of using the top-\textit{K} method to eliminate nodes, we perform the weighted summation operation over all nodes based on self-attention values to ensure that each node in the SPG contributes to the result. Therefore, the feature vector of the graph $g^G$ can be calculated in (\ref{eq4}). $\alpha_i^{G}$ is the attention weight corresponding to node $v_i^{G}$, which is obtained by normalizing the attention scores of all nodes in the graph $g^G$ by the softmax function. $V_{G}=\{v_1^{G},\dots,v_i^{G},\dots,v_{\vert V_{G}\vert}^{G}\}$ is the set of nodes of the graph $g^G$ and $z_i^{G}$ is the attention score corresponding to node $v_i^{G}$. $S_{G}\in R^{(L+1)\cdot z}$ is the feature vector  of the graph $g^G$.

\textbf{Subgraph-level Attention mechanism:} After obtaining the vector representations of the SPG, $CDG_{SPG}$, $DDG_{SPG}$ and $FCDG_{SPG}$, we propose the subgraph-level attention mechanism to obtain the feature vector of the combined three types of subgraphs. In our approach, the attention of a subgraph of SPG indicates the importance of the subgraph to the SPG, which can be calculated in (\ref{eq5}).
\vspace{-0.1cm}
\begin{equation}
\begin{aligned}
&r_{sub}=S_{sub}^TW_rS_{SPG}\\
&sub\in \{CDG_{SPG},DDG_{SPG},FCDG_{SPG}\}
\end{aligned}\label{eq5}
\vspace{-0.1cm}
\end{equation}

Where $sub$ denotes the type related to the subgraph. $r_{sub}\in R$ and $S_{sub}^T\in R^{(L+1)\cdot z}$ represent the attention score and the feature vector of the graph $sub$ respectively. $W_r\in R^{(L+1)z\times (L+1)z}$ is the matrix of weights to be learned. $S_{SPG}$ denotes the feature vector  of the $g^{SPG}$. Then, the attention weights of the subgraphs are obtained by the softmax function. Finally, the subgraph feature vectors are lineraly combined through a weighted sum which is calculated as follows:
\begin{equation}
\vspace{-0.1cm}
\begin{aligned}
&\beta_{sub}=\frac{exp(r_{sub})}{\sum_{sub}exp(r_{sub})}\\
&S_{AS}=\sum_{sub}\beta_{sub}\cdot S_{sub}
\end{aligned}\label{eq6}
\end{equation}

Where $\beta_{sub}$ is the attention weight corresponding to the graph $sub$, $S_{AS}$ is the feature vector of all subgraphs.

\subsection{The Classifier Network}
At last, we concatenate the $S_{SUB}$ and the $S_{SPG}$ obtained in Section \ref{IVD}, and then feed them directly into the classifier network for vulnerability detection. To achieve end-to-end learning, the classifier network in our approach is a single fully connected layer with softmax outputs. The formula for the classifier network is (\ref{eq7}).
\begin{equation}
\begin{aligned}
p(y|g^{SPG})=softmax(W_{CN}(S_{SPG}||S_{AS})+b_{CN})
\end{aligned}\label{eq7}
\end{equation}

Where $y\in \{0,1\}$ is the class label, and 1 is a vulnerable sample, 0 is a non-vulnerable sample. Both $W_{CN}\in R^{2(L+1)z\times 2}$ and $b_{CN}\in R^2$ are the learnable parameters of the classifier network. $p(y|g^{SPG})$ is the output of the classifier network. Finally, we use the widely used cross-entropy function \cite{bishop2006pattern} as a loss function to train the model.
\begin{equation}
\begin{aligned}
L=-\sum_{g^{SPG}\in Train}\sum_{y\in \{0,1\}}\Tilde{p}(y|g^{SPG})log(p(y|g^{SPG}))
\end{aligned}\label{eq8}
\end{equation}

Where $Train$ is the training datasets, $\Tilde{p}(y|g^{SPG})$ is the ground truth, i.e., $\Tilde{p}(y|g^{SPG})=1$ if $g^{SPG}$ contains vulnerabilities and $\Tilde{p}(y|g^{SPG})=0$ otherwise. 

\section{EVALUATION}

\subsection{Evaluation Setup}

\textbf{Research Questions:} Our evaluation is designed to answer the following research questions.

RQ1: How effective is VulSPG when compared with state-of-the-art vulnerability detection methods at slice level?

RQ2: How effective is VulSPG when compared with state-of-the-art vulnerability detection methods at function level? 

RQ3: How effective is SPG in vulnerability detection tasks when compared with other graph-based code representations?

RQ4: Is our triple attention mechanism beneficial to vulnerability detection?

\begin{table}[b]
\vspace{-0.6cm}
\caption{The number of SPGs, vulnerable SPGs, non-vulnerable
SPGs from the programs of FFmpeg and QEMU}
\vspace{-0.4cm}
\begin{center}
\begin{tabular}{|c|p{19pt}|p{19pt}|p{0.9cm}|p{19pt}|p{19pt}|p{0.9cm}|}
\hline
\textbf{Kind of}&\multicolumn{3}{|c|}{\textbf{SPGs of FFmpeg}} &\multicolumn{3}{|c|}{\textbf{SPGs of Qemu}}\\
\cline{2-4} \cline{5-7}
\textbf{SyVCs} & \textbf{\#SPGs} & \textbf{\#Vul.-SPGs} & \textbf{{\#Non-vul.SPGs}} & \textbf{\#SPGs} & \textbf{\#Vul.-SPGs} & \textbf{\#Non-vul.SPGs} \\
\hline
FC-kind&8,388&2,874&5,514&1,2258&3,253&9,005  \\
\hline
AU-kind&18,914&6,471&12,443	&7,861	&1,623	&6,238  \\
\hline
PU-kind&32,238&13,120&19,118&44,705&12,897&31,808\\
\hline
AE-kind&13,387&5,866&7,521	&5,677	&1,761	&3,916\\
\hline
FP-kind	&30,636	&12,311	&18,325	&39,763	&10,492	&29,271\\
\hline
FR-kind	&24,510	&8,154	&16,356	&26,195	&5,830	&20,365\\
\hline
Total	&128,073	&48,796	&79,277	&136,459	&35,856	&100,603\\
\hline
\end{tabular}
\label{tab1}
\end{center}
\vspace{-0.1cm}
\end{table}
\textbf{Vulnerability dataset:} We conduct comparative experiments on two datasets at the slice level and the function level to evaluate the benefits of VulSPG. The first dataset is provided by Li et al.\cite{li2021sysevr} and collected from the National Vulnerability Database (NVD) \cite{NVDnvd} and the Software Assurance Reference Dataset (SARD) \cite{SARDsard}. There are 15,591 programs in this dataset, of which 14,000 are from SARD and contain 13,906 vulnerability programs, and the remaining 1,591 programs from NVD with 874 vulnerability programs. Most of the programs in this dataset are synthetic, which is not suitable for evaluating the performance of the model in practical applications. In addition, this dataset contains interference information (e.g., keyword "Good" and "Bad", etc.), so we peformed a normalization on source code in this dataset and used the same normalization method as in \cite{li2021sysevr} for a fair comparison. The second dataset provided by Zhou et al. \cite{zhou2019devign} contains real-world programs extracted from two C/C++ open-source projects(i.e. FFmpeg and QEMU), including 17,549 programs from QEMU, of which 7,479 are vulnerable, and 9,769 programs from FFmpeg, 4,981 of which are vulnerable. The number of SPGs extracted  from these two projects is shown in Table \ref{tab1}.

By introducing two new types of SyVCs (i.e. FP-kind and FR-kind), the vulnerability coverage of SPGs on two datasets increased by 3.1\% and 10.0\%, respectively.

\textbf{labeling SPG:} For SARD programs in the first dataset, if the statement node contains at least one vulnerability in the generated SPG, it is labeled as 1, otherwise it is 0. Unlike the SARD programs, the programs from NVD, FFmpeg and QEMU have no explicit vulnerable lines of code. Therefore, we can only generate labels for SPGs through the diff files, which are divided into three steps. First, for all SPGs generated by non-vulnerable programs, we label them as 0. Second, for deleted lines in the diff file, we treat them as vulnerability codes. That is, if there is a corresponding deleted line in the statement nodes of SPG, the SPG will be labeled as 1. Third, for programs that have no deleted lines in the diff files but are vulnerable, we treat its overall structure as vulnerable, that is, all SPGs generated by these programs are labeled as 1. Finally, we manually review SPGs labeled as vulnerable by security experts to ensure proper labeling.

\textbf{Baseline methods:} We select slice-based approaches (VulDeePecker \cite{li2018vuldeepecker} and SySeVR \cite{li2021sysevr}) and graph-based approaches (Devign \cite{zhou2019devign} and FUNDED \cite{wang2020combining}) as comparisons to evaluate the effectiveness of the proposed methods.  VulDeePecker extracts code segments called code gadgets to represent the programs and then uses BiLSTM to learn the vulnerability patterns. SySeVR generates the code segments named SeVCs by extracting four types of SyVCs and then uses BiGRU to learn the vulnerability patterns. Devign learns vulnerability patterns from the code composite graphs using GGNN with the CONV module for vulnerability detection. FUNDED is similar to Devign, but the proposed representation of the codes is more complex and contains more syntactic and structural information and the improved GGNN is verified to be more suitable for learning multiple relationships of complex graphs.

\textbf{Evaluation Configuration:} We implement the proposed GNN-based model with the triple attention mechanism using Pytorch v1.6 \cite{Pytorch2021} and DGL\cite{DGL2021}. Our experiments were run on a machine with an Intel Core 2.6 GHz CPU and an NVIDIA 2080Ti GPU. Taking into account the computational efficiency of VulSPG and the results of parameter adjustment during the experiments, we make the following parameter settings. The dimensionality of initial node vector is 64. In the graph encoding phase, we set the dimensionality of hidden states as 64, and the number of hidden state layers as 2. Therefore, the dimensionality of final vector of a SPG is 384. We train our model using the Adam optimizer with a learning rate of 0.001 and 200-epoch patience for early stopping. All baseline approaches are configured with the same settings as reported in their original papers\cite{li2018vuldeepecker} \cite{li2021sysevr} \cite{zhou2019devign} \cite{wang2020combining}. 

\textbf{Evaluation Metrics:} We use six widely used metrics to evaluate the performance of VulSPG and other baseline approaches.
\textbf{Accuracy (A)}. The proportion of all test cases that are classified correctly.
\textbf{Precision (P)}. The proportion of correctly predicted vulnerability samples to those predicted to be vulnerable.
\textbf{Recall (R)}. The proportion of correctly predicted vulnerability samples to all vulnerable samples.
\textbf{False-positive rate (FPR)}. The proportion of false-positive samples among all non-vulnerable samples.
\textbf{False-negative rate (FNR)}. The proportion of false-negative samples among all vulnerable samples.
\textbf{F1-measure (F1)}. F1 assesses the overall effect by considering precision and recall.

\subsection{Evaluation Results}

\textbf{Experiments for answering RQ1:} To investigate the answer to Q1, we conducted vulnerability detection experiments on the first dataset at the slice level. For comparison, we chose three static vulnerability detection tools: checkmark \cite{checkmarx2021}, Flawfinder \cite{flawfinder2021}, and RATS \cite{RATs2021} and two slice-based methods: VulDeePecker \cite{li2018vuldeepecker} and SySeVR \cite{li2021sysevr}. To reduce the cost and increase the credibility of the experiment, we use the experimental results from the original paper \cite{li2021sysevr}. Following the experimental configurations in \cite{li2021sysevr}, we randomly select 80\% of the programs in the first dataset as the training set and the rest 20\% as the testing set. Specifically, we extracted 639,703 SPGs corresponding to the six kinds of SyVCs from 12,473 programs to train the VulSPG model, and 158,084 SPGs extracted from 3,118 programs to test the model.
\begin{table}[t]
\caption{Comparing our approach with static vulnerability detection tools and
state-of-the-art slice-based methods at the slice level (metrics unit:\%)}
\vspace{-0.4cm}
\begin{center}
\begin{tabular}{|c|c|c|c|c|c|c|}
\hline
\textbf{Method}&\textbf{FPR}&\textbf{FNR}&\textbf{A}&\textbf{P}&\textbf{R}&\textbf{F1} \\
\hline 
Flawfinder&21.6 &70.4 &69.8 &22.8 &29.6 &25.7 \\
\hline
RATS	&21.5	&85.3	&67.2	&12.8	&14.7 &13.7 \\
\hline
Checkmarx	&20.8	&56.8	&72.9	&30.9&43.4	&36.1 \\
\hline
VulDeePecker	&2.5	&41.8	&92.2	&78&58.2	&66.6 \\
\hline
SySeVR-BGRU	&\textbf{1.4}	&5.6	&\textbf{98.0}	&92.6&94.4	&90.5 \\
\hline
\textbf{Our approach}	&2	&\textbf{5.3}	&97.1	&\textbf{93.9}&\textbf{94.7}	&\textbf{94.3} \\
\hline
\end{tabular}
\label{tab2}
\end{center}
\vspace{-0.6cm}
\end{table}

\begin{table*}[t]
\caption{Comparisons between our method and other methods on the second dataset at the function level }
\vspace{-0.4cm}
\begin{center}
\begin{tabular}{|c|c|c|c|c|c|c|c|c|c|c|c|}
\hline
\textbf{Project}&\textbf{Method}&\textbf{TP}&\textbf{TN}&\textbf{FP}&\textbf{FN}&\textbf{FPR(\%)}&\textbf{FNR(\%)}&\textbf{A(\%)}&\textbf{P(\%)}&\textbf{R(\%)}&\textbf{F1(\%)} \\
\hline
&VulDeePecker	&156	&193	&89	&163	&\textbf{31.6}	&51.1	&58.1	&63.7&48.9	&55.3 \\
\cline {2-12}
&SySeVR(4-type)	&655	&650	&396	&399	&37.9	&37.8	&62.1	&62.3&62.2	&62.2 \\
\cline {2-12}
FFmpeg&SySeVR(6-type)	&743	&750	&415	&409	&35.6	&35.5	&64.4	&\textbf{64.2}&64.5	&64.3 \\
\cline {2-12}
&Devign	&781	&512	&653	&371	&56.1	&32.2	&55.8	&54.5&67.8	&60.4 \\
\cline{2-12}
&FUNDED &864 &397 &768 &288 &65.9 &25.0 &54.4 &52.9 &75.0 &62.1 \\
\cline{2-12}
&\textbf{Our method}&947&551&614&205&52.7&\textbf{17.8}&\textbf{64.7}&60.7&\textbf{82.2}&\textbf{69.8} \\
\hline
&VulDeePecker	&145	&340	&73	&296	&17.7	&67.1	&56.8	&66.5&32.9	&44.0 \\
\cline{2-12}
&SySeVR(4-type)	&636	&1,534	&362	&850	&19.0	&57.2	&64.2	&63.7&42.8	&51.2 \\
\cline{2-12}
QEMU&SySeVR(6-type)	&770	&2,030	&350	&890	&\textbf{14.7}	&53.6	&\textbf{69.3}	&\textbf{68.8} &46.4	&55.4 \\
\cline{2-12}
&Devign	&715	&1,684	&696	&945	&29.2	&56.9	&59.4	&50.7&43.1	&46.6 \\
\cline{2-12}
&FUNDED &744&1,693&687&916&24.7&55.2&62.8&55.9&44.8&49.7 \\
\cline{2-12}
&\textbf{Our method}&1,010&1,749&631&650&26.5&\textbf{39.2}&68.3&61.6&\textbf{60.8} &\textbf{61.2}\\
\hline
\end{tabular}
\label{tab3}
\end{center}
\vspace{-0.6cm}
\end{table*}

The experimental results are summarized in Table \ref{tab2}. First, all slice-based methods perform better than the static vulnerability detection tools. Second, our approach outperforms the other slice-based methods. VulSPG is 4.9\% more accurate and the F1-measure is 27.7\% higher than VulDeePecker. Compared to SySeVR, VulSPG is still superior in pecision, recall and F1-measure, although the accuracy is slightly lower. \textbf{\emph{In summary, VulSPG is more effective than static vulnerability detection tools and state-of-the-art slice-based approaches on slice-level vulnerability detection tasks. }}

Note that for RQ2, RQ3, and RQ4, we focus on performing experiments on the second dataset. The main reasons are as follows. First, the second dataset consists of vulnerable codes from real-world projects, while most of the samples in the first dataset are synthetic codes. Second, the second dataset, generated on function-level, is more appropriate for comparing our approach with the state-of-the-art graph-based methods.

\textbf{Experiments for answering RQ2:} This experiment aim to demonstrate the effectiveness of the VulSPG at the function-level granularity. Note that SySeVR (4-type) denotes the baseline method based on four types of SyVCs (FC-kind, AU-kind, PU-kind, AE-kind) proposed in their original paper, and SySeVR (6-type) indicates the addition of our proposed two SyVCs (FR-kind and FP-kind). Referring to the experimental setup in \cite{zhou2019devign}, we randomly divide the samples into two groups at a ratio of 75\%:25\%. Unlike other graph-based program representations, SPGs are generated by using program slicing technique according to a series of criterions (i.e., SyVCs) extracted from the program. Therefore, the number of SPGs generated in the dataset is much larger than the number of programs. To make a fair comparison with graph-based methods Devign and FUNDED, We use the following methods to detect program vulnerabilities. For each program in the testing dataset, a program is predicted to be vulnerable if at least one of its SPGs is predicted to be vulnerable. A program is predicted to be non-vulnerable if all its SPGs are predicted to be non-vulnerable. Slice-based methods VulDeePecker and SySeVR also use the above approach for vulnerability detection. The experimental results are presented in Table \ref{tab3}.

The experimental results indicate that our proposed VulSPG outperforms other state-of-the-art methods on vulnerability detection task at the function-level granularity. In the slice-based approaches, SySeVR performs better than VulDeePecker. The primary reason is that VulDeePecker only detects vulnerabilities related to insecure library functions, and thus cannot cover vulnerable programs in the testing dataset that do not contain insecure functions when generating code segments. Similarily, SySeVR (6-type) covers more types of vulnerabilities than SySeVR (4-type) and, thus, has better performance under the same testing dataset. 
In the graph-based approaches, FUNDED slightly outperforms the Devign in some of the metrics. Although both of them use GGNNs, FUNDED improves upon GGNN to make it more suitable for modeling multiple code relationships extracted from the source code. However, unexpectedly, the graph-based methods (i.e., Devign and FUNDED) perform worse than the slice-based methods (e.g., SySeVR (4-type)). The reason is in two folds: (1) As we detailed in Section~\ref{IIA}, the graph-based representations of the code used by Devign and FUNDED contain a lot of nodes that are not relevant to any vulnerability, which greatly hinders the GGNN to learn effective vulnerable features. 
(2) For the FUNDED and Devign methods, the token embedding within a statement is simply averaged or fed into the linear layer to obtain initial node representation, which fails to account for the dependencies between tokens in the statement nodes. \textbf{\emph{In summary, VulSPG is more effective than the state-of-the-art slice-based and graph-based methods on the function-level vulnerability detection task.}}

\textbf{Experiments for answering RQ3:} In this experiment, AST, CFG, PDG, and CPG, are selected as the baselines of program representation. R-GCN and GGNN are selected as the baselines of vulnerability detection model. In the model readout phase, we uniformly use the multilayer perceptron (MLP) for graph-level classification to mitigate the impact of the model on vulnerability detection performance. The detection results are presented in Table \ref{tab4}.
\begin{table}[b]
\vspace{-0.7cm}
\caption{Vulnerability detection results of different graph-based program representations (metrics unit:\%)}
\vspace{-0.4cm}
\begin{center}
\begin{tabular}{|c|c|c|c|c|c|}
\hline
\textbf{project}&\textbf{represe-}&\multicolumn{2}{|c|}{\textbf{RGCN}}&\multicolumn{2}{|c|}{\textbf{GGNN}} \\
\cline {3-6}
&\textbf{tations}&\textbf{A}&\textbf{F1}&\textbf{A}&\textbf{F1} \\
\hline 
&AST&57.6&59.6&55.8&57.3 \\
\cline {2-6}
&CFG&60.4	&59.9&56.0	&59.4 \\
\cline {2-6}
FFmpeg&PDG&60.0	&61.2&57.2&60.0 \\
\cline {2-6}
&CPG&54.5	&60.9&53.5	&58.7 \\
\cline {2-6}
&\textbf{SPG}&\textbf{62.1}	&\textbf{64.7}&\textbf{58.9}	&\textbf{61.5} \\
\hline
&AST&57.2	&49.3&57.9	&42.8 \\
\cline {2-6}
&CFG&59.3	&53.1&58.3	&46.1 \\
\cline {2-6}
QEMU&PDG&58.8	&54.0&58.2	&47.8 \\
\cline {2-6}
&CPG&57.8	&51.6&57.3	&45.7 \\
\cline {2-6}
&\textbf{SPG}&\textbf{64.5}	&\textbf{58.6}&\textbf{62.9}	&\textbf{57.6} \\
\hline
\end{tabular}
\label{tab4}
\end{center}
\vspace{-0.15cm}
\end{table}

As shown in Table \ref{tab4}, the experimental results for CFG and PDG are slightly better than those for AST. In fact, although AST contains more syntax information than CFG and PDG, it also contains more nodes and edges that are not relevant to vulnerabilities. Simlar to AST, although CPG contains rich information to allow detailed representation of various types of vulnerability patterns, the large amount of redundant information unrelated to vulnerabilities it contains also hinders vulnerability detection, which eventually leads to its failure to obtain better performance of vulnerability detection. The proposed SPG combines slicing technique and graph-based representation to remove a large amount of redundant information while preserving as much structural and semantic information about the vulnerability as possible. Therefore, it achieves the best results in the experiment. The GGNN model does not perform as well as expected, and we found that the GGNN model is prone to over-smoothness in the learning process. GGNN cannot learn the property information of different types of directed edges in SPG like R-GCN, which also explains the performance drop. \textbf{\emph{In summary, our proposed SPG are more effective than other graph-based code representations in vulnerability detection.}}

\textbf{Experiments for answering RQ4:} To demonstrate the effectiveness of the triple attention mechanism in the VulSPG, we use R-GCN but different readout methods to learn vulnerability patterns in SPG. Specifically, We choose MLP and Conv as baselines, where Conv is the readout method of Devign that allows more effective feature extraction. Next, we conduct ablation experiments targeting the triple attention mechanism. We use the full triple attention mechanism as a comparison, removing the token-level, the node-level, and the subgraph-level attention mechanism, respectively. The experimental results are shown in Table \ref{tab5}.
\begin{table}[b]
\vspace{-0.6cm}
\caption{Ablation experiments targeting the triple attention mechanism (metrics unit:\%)}
\vspace{-0.4cm}
\begin{center}
\begin{tabular}{|c|c|c|c|c|}
\hline
\textbf{}&\multicolumn{2}{|c|}{\textbf{FFmpeg}}&\multicolumn{2}{|c|}{\textbf{QEMU}}\\
\cline{2-5}
\textbf{Model}&\textbf{A}&\textbf{F1}&\textbf{A}&\textbf{F1}\\
\hline
RGCN+MLP	&62.1	&64.7	&64.5	&58.6\\
\hline
RGCN+Conv	&60.2	&66.5	&66.3	&58.5\\
\hline
RGCN+token+subgraph	&63.4	&67.1	&67.8	&59.1\\
\hline
RGCN+token+node	&63.4	&65.7	&67.2	&57.3 \\
\hline
RGCN+node+subgraph	&61.7	&66.1	&67.5	&56.8 \\
\hline
\textbf{RGCN+triple attention}&\textbf{64.7}&\textbf{69.8}&\textbf{68.3}&\textbf{61.2}\\
\hline
\end{tabular}
\label{tab5}
\end{center}
\end{table}

The experimental results show that the triple attention mechanism improves the graph classification performance compared with the MLP and Conv models. Indeed, focusing on the structural and semantic features of the code through the attention mechanism is more suitable for vulnerability detection than collecting all node information equally. In the ablation experiment, we find that removing subgraph-level attention mechanism has a greater impact on performance, suggesting that subgraph-level attention mechanism is most important in the triple attention mechanism. The token-level attention mechanism is used in the process of SPG node embedding, and from the experimental results, it plays an equally important role. This indicates that the token-level attention mechanism extracts dependency information between tokens within a statement, thus improving the performance of vulnerability detection. To our surprise, the node-level attention mechanism, while playing a role, is not as important as we expected. The above experimental results demonstrates to some extent that accurate identification of vulnerability patterns requires the ability to consider the combination of semantic, structural, and contextual information about the vulnerability code.

\textbf{\emph{In summary, the triple attention mechanism achieves higher vulnerability detection capability and where token-level, node-level, subgraph-level attention mechanisms are all effective}.}

\subsection{Discussion and Limitation}
\textbf{Inter-procedural analysis} In our experiments, the Devign’s dataset is not provided with a complete program and contains only function-level code, thus we did not perform the inter-procedural analysis on this dataset. We find that despite the absence of inter-procedural analysis, VulSPG still performs well on Devign's dataset. In addition, using the original same dataset also allows us to make a fairer comparison with Devign. Our future work will consist in verifying the performance of the inter-procedural analysis in real-world projects.

\textbf{Bias in the experimental results.} In the experiments of RQ2, the results of our implementation of Devign differ from those reported in the original paper \cite{zhou2019devign}. This is because in the experiments we used all the programs in the second dataset and did not filter out programs with too many tokens, as in the original paper. Using the FFmpeg project as an example, Devign use 6,716 programs when performing the experiments, while we use all 9,769 programs in the experiments. This leads to more difficult identification of vulnerabilities and worse experimental results. In addition to using different datasets, it is difficult to reproduce the code and optimize the experimental parameters in the same way as the original paper because the original authors did not open the implemented source code.


\textbf{Limitations.} \textbf{First}, the VulSPG currently focuses on detecting vulnerabilities in the source code of C/C++ programs. We plan to extend this method to other languages in future work. \textbf{Second}, although we have added two kinds of SyVCs, there are still some vulnerability codes that are not covered by our generated SPGs. \textbf{Third}, we only consider the most commonly used vulnerability detection method GGNN as the comparison of graph coding. We will compare more types of GNN in our future work.

\section{RELATED WORK}

Software vulnerability detection relies on program analysis techniques, which include static analysis\cite{pistoia2007survey}, dynamic analysis\cite{godefroid2008automated}, and hybrid analysis\cite{shar2014web}. Our work builds upon the foundations of static program analysis techniques.

\textbf{Classic methods:} Early static vulnerability detection methods are based on rules developed manually by security experts\cite{mahmood2018evaluation} \cite{engler2001bugs} \cite{RoughAuditing}. However, these methods rely on the knowledge and experience of experts and it is difficult to develop quality rules. Symbolic execution \cite{cadar2008klee} \cite{cadar2013symbolic} mitigates this by analyzing the control flow of the program, but also suffers from problems such as path explosion and difficulty in resolving constraints.

\textbf{Machine leaning based methods:} Machine learning (ML) based vulnerability detection methods offer alternative solutions for automated and more efficient vulnerability detection. 
Using manually defined features extracted from source code, such as function/library calls\cite{neuhaus2007predicting}, software complexity metrics \cite{chowdhury2011using} \cite{shin2008empirical} \cite{zimmermann2010searching} and developer activity \cite{shin2010evaluating}, ML-based approaches learn potential or abstract vulnerability patterns by machine learning algorithms. The performance of these machine learning methods strongly depends on the quality of the features extracted from the data (i.e. the discriminative and expressive power of the features) \cite{ghaffarian2017software}. 

\textbf{Deep Learning based methods:} DL-based detection methods are able to extract features of vulnerabilities and automatically learn patterns of vulnerable code, which can significantly reduce time and labor costs as they do not require human-defined features \cite{lin2020software} \cite{shen2020survey} \cite{singh2020applying}. We classify the representative DL-based vulnerability detection methods into slice-based methods (e.g., VulDeePecker \cite{li2018vuldeepecker} and SySeVR \cite{li2021sysevr} ) and graph-based methods (e.g., Devign \cite{zhou2019devign} and FUNDED \cite{wang2020combining}). Slice-based methods retain only information that may be relevant to the vulnerability, but still treat the code as sequential tokens and often ignore structure of the code. The graph-based methods take into account the structure and logic of the code. However, these code representations also contain too much redundant information unrelated to the vulnerabilities, which interferes with the GNN learning the vulnerability patterns from the vulnerable code.

\section{CONCLUSION}

We propose VulSPG, which contains a code representation called SPG for vulnerability detection tasks, and a vulnerability detection method using R-GCN with a triple attention mechanism. We evaluate VulSPG on a large-scale slice-level vulnerability dataset and a function-level real-world dataset, respectively. The experimental results demonstrate that VulSPG is superior to state-of-the-art approaches.

\section*{Acknowledgment}

This work was supported by the National Natural Science Foundation of China (Grant Nos.61672191)


\bibliographystyle{IEEEtran}
\bibliography{reference}

\end{document}